\documentstyle[epsf,12pt]{article}

\makeatletter

 \@addtoreset{equation}{section}
\makeatother

\begin{document}
\topmargin 0pt
\oddsidemargin 5mm

\newcommand {\beq}{\begin{eqnarray}}
\newcommand {\eeq}{\end{eqnarray}}
\newcommand {\non}{\nonumber\\}
\newcommand {\eq}[1]{\label {eq.#1}}
\newcommand {\defeq}{\stackrel{\rm def}{=}}
\newcommand {\gto}{\stackrel{g}{\to}}
\newcommand {\hto}{\stackrel{h}{\to}}

\newcommand {\1}[1]{\frac{1}{#1}}
\newcommand {\2}[1]{\frac{i}{#1}}

\newcommand {\th}{\theta}
\newcommand {\thb}{\bar{\theta}}
\newcommand {\ps}{\psi}
\newcommand {\psb}{\bar{\psi}}

\newcommand {\ph}{\varphi}
\newcommand {\phs}[1]{\varphi^{*#1}}

\newcommand {\sig}{\sigma}
\newcommand {\sigb}{\bar{\sigma}}
\newcommand {\Ph}{\Phi}
\newcommand {\Phd}{\Phi^{\dagger}}
\newcommand {\Sig}{\Sigma}
\newcommand {\Phm}{{\mit\Phi}}

\newcommand {\eps}{\varepsilon}
\newcommand {\del}{\partial}
\newcommand {\dagg}{^{\dagger}}
\newcommand {\pri}{^{\prime}}
\newcommand {\prip}{^{\prime\prime}}
\newcommand {\pripp}{^{\prime\prime\prime}}
\newcommand {\prippp}{^{\prime\prime\prime\prime}}
\newcommand {\delb}{\bar{\partial}}

\newcommand {\zb}{\bar{z}}
\newcommand {\mub}{\bar{\mu}}
\newcommand {\nub}{\bar{\nu}}
\newcommand {\lam}{\lambda}
\newcommand {\lamb}{\bar{\lambda}}
\newcommand {\kap}{\kappa}
\newcommand {\kapb}{\bar{\kappa}}
\newcommand {\xib}{\bar{\xi}}
\newcommand {\Ga}{\Gamma}
\newcommand {\rhob}{\bar{\rho}}
\newcommand {\etab}{\bar{\eta}}

\newcommand {\zbasis}[1]{\del/\del z^{#1}}
\newcommand {\zbbasis}[1]{\del/\del \bar{z}^{#1}}
\newcommand{\vs}[1]{\vspace{#1 mm}}
\newcommand{\hs}[1]{\hspace{#1 mm}}

\begin{titlepage}

\begin{flushright}
OU-HET 269\\
KEK-Prep.97-93\\
KEK-TH-524\\
hep-th/9706219\\
\end{flushright}
\bigskip
\bigskip

\begin{center}
{\LARGE\bf
Low Energy Theorems
in N=1 Supersymmetric Theory
}
\vs{10}

{\renewcommand{\thefootnote}{\fnsymbol{footnote}}
{\large\bf Kiyoshi Higashijima,$^1$\footnote{higashij@phys.wani.osaka-u.ac.jp.}
Muneto Nitta,$^{1,2}$\footnote{
 nittam@theory.kek.jp, nitta@phys.wani.osaka-u.ac.jp.}
Kazutoshi Ohta$^1$\footnote{kohta@funpth.phys.sci.osaka-u.ac.jp.
Supported in part by the JSPS Research Fellowships.}\\
and \\
Nobuyoshi Ohta$^1$\footnote{ohta@phys.wani.osaka-u.ac.jp.}\\
}}
\setcounter{footnote}{0}
\bigskip

{\small \it
$^1$Department of Physics,
Graduate School of Science, Osaka University,\\
Toyonaka, Osaka 560, Japan\\
$^2$Theory Division,
Institute of Particle and Nuclear Studies, KEK,\\
Tsukuba, Ibaraki 305, Japan
}
\end{center}
\bigskip

\begin{abstract}
In $N=1$ supersymmetric theories, quasi Nambu-Goldstone (QNG) bosons
appear in addition to ordinary Nambu-Goldstone (NG) bosons
when the global symmetry $G$ breaks down spontaneously.
We investigate two-body scattering amplitudes of these bosons
in the low-energy effective Lagrangian formalism.
They are expressed by the curvature of K\"{a}hler manifold.
The scattering amplitudes of QNG bosons are shown to coincide with
those of NG bosons though the effective Lagrangian contains an arbitrary
function, and those with odd number of QNG bosons all vanish.
\end{abstract}

\end{titlepage}

\section{Introduction}

When the global symmetry $G$ breaks down spontaneously to its subgroup $H$,
there are massless Nambu-Goldstone (NG) bosons, as many as the broken
generators. The low energy theorems say that
the low-energy scattering amplitudes of NG bosons are
determined only by the symmetries $G$ and $H$.
If we integrate out the massive particles,
we obtain the low-energy effective Lagrangian
which reproduces the scattering amplitudes of NG bosons.

Callan, Coleman, Wess and Zumino (CCWZ) constructed
such a low-energy effective Lagrangian
as a nonlinear sigma model whose target manifold is
the coset $G/H$ parametrized by NG bosons~\cite{CCWZ}.
We can calculate the low-energy scattering amplitudes of
NG bosons by summing up tree graphs. The target manifold $G/H$
is homogeneous and compact and has $G$-invariant metric.
The broken symmetry $G$ is realized nonlinearly while
the unbroken symmetry $H$ is represented linearly.

In theories with unbroken $N=1$ supersymmetry, there come out some extra
massless quasi Nambu-Goldstone (QNG) bosons besides the ordinary NG bosons.
The target manifold of the $N=1$ nonlinear sigma model in four dimensions
must be K\"{a}hler~\cite{Zu} with $G$-isometry and is parametrized by the
NG and QNG bosons. General methods to construct K\"{a}hler potential which
determines the effective Lagrangian are given by Bando, Kuramoto, Maskawa
and Uehara (BKMU) in their celebrated papers~\cite{BKMU}.

In supersymmetric theories, the global symmetry $G$ is spontaneously broken
by the F-term superpotential which is holomorphic function of the chiral
superfields. As a result, the symmetry $G$ is enhanced to its complexification
$G^{\bf C}$, and the isotropy group $H$ is extended to $\hat{H}$ which
contains the complexification $H^{\bf C}$ of $H$~\cite{Le,BKMU}.
BKMU constructed the following three
types of K\"{a}hler potentials for the target manifold $G^{\bf C}/\hat H$.
The A- and C-type Lagrangians contain arbitrary functions of the $G$-invariant
quantities. The B-type Lagrangian is a generalization of the Zumino's
Lagrangian~\cite{Zu}, and does not have such arbitrariness.

When there is no QNG boson, it has been proved by Itoh, Kugo and Kunitomo
(IKK)~\cite{IKK} that the most general Lagrangian is uniquely determined
by the B-type Lagrangian. In this case the target manifold $G^{\bf C}/\hat H$
is isomorphic to $G/H$, and the unbroken group $H$ must be the centralizer of
a torus. This means that the homogeneous manifold $G/H$ should be K\"{a}hler.
This condition restricts the breaking pattern and is used to
classify all the compact homogeneous K\"{a}hler $G/H$ Lagrangian~\cite{IKK}.

{}For the low-energy effective Lagrangian of linear field theory origin,
there must be at least one QNG boson~\cite{Le,KS},
so that we cannot use the IKK construction.
Instead, we must use the A- or C-type Lagrangians of BKMU.\footnote{
In this paper, we use only the A-type Lagrangian,
because little is known about the C-type Lagrangian of BKMU.}
As we already remarked, these contain arbitrary functions due to the
existence of QNG bosons. The origin of these functions
is understood geometrically as follows.

The target manifold $G^{\bf C}/\hat H$ is non-compact and
the QNG bosons correspond to the non-compact directions~\cite{Le,BL,KS}.
Since the symmetry of the Lagrangian is still $G$ but not $G^{\bf C}$,
the metric in the QNG directions is not determined by the symmetry,
whereas that in the NG directions is determined by the low energy theorems.
The arbitrary function thus describes the shape of non-compact directions of
the target manifold~\cite{BL,KS}.

The arbitrariness discussed above suggests that the scattering amplitudes
of the QNG bosons will not be determined by symmetries and there may not
be any low energy theorem. (PCAC relation in supersymmetric theory
has been discussed in ref.~\cite{Le2}.) In this paper, we investigate the
low energy theorems for the scattering amplitudes of NG and QNG bosons,
and show that the scattering amplitudes of only the QNG bosons actually
coincide with those of NG bosons, in spite of the arbitrariness.
The arbitrariness shows up in the scattering amplitudes between NG and QNG
bosons. We also confirm that the amplitudes of NG in supersymmetric theories
are the same as those in the non-supersymmetric theory.

In sect.~2, we review the CCWZ bosonic theories and the low energy theorems.
In sect.~3, we proceed to the supersymmetric theories and derive the low
energy theorems. Sect.~4 is devoted to conclusions and discussions.

\section{Scattering amplitudes of
NG bosons in non-supersymmetric theory}

In this section, we start with the bosonic theories.
The most general low-energy effective Lagrangian is expressed as
\beq
{\cal L}(\phi)
 = \1{2} g_{\alpha\beta}(\phi)
   \del_{\mu} \phi^{\alpha} \del^{\mu} \phi^{\beta},
 \label{general Lagrangian}
\eeq
where we use $\alpha, \beta, \cdots$ for curved indices on the target
manifold and $\mu$ for spacetime coordinates. This is a nonlinear sigma
model on a general target manifold with a metric $g_{\alpha\beta}(\phi)$.
The scattering amplitudes calculated from this Lagrangian
are invariant under the field redefinitions which may be interpreted as
the general coordinate transformations on the target manifold.
Using this freedom, we may choose the most convenient Riemann normal
coordinate (figure 1). It is the coordinate spanned by the geodesic
lines $\lambda^i(t) (0\leq t \leq 1)$ starting from a point
determined by the vacuum expectation value.\footnote{
We use $i, j, \cdots$ for the flat tangent space indices,
which coincide with the indices of broken generators at $\phi=\phi_0$.}

\begin{figure}
 \epsfxsize=6.5cm
 \centerline{\epsfbox{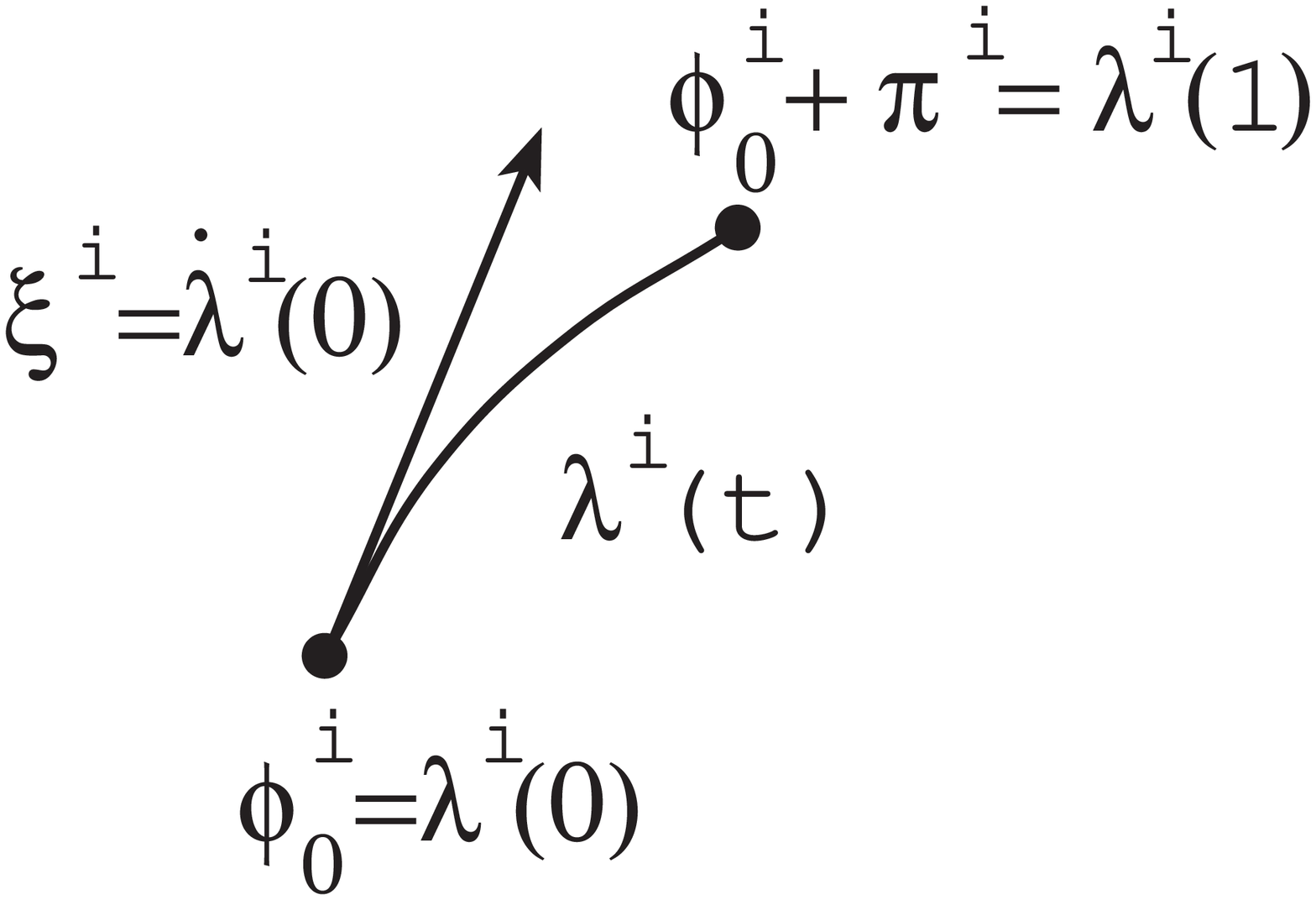}}
 \centerline{\bf Figure\,1.\,Riemann normal coordinate}
\end{figure}
To compute scattering amplitudes, we expand the effective Lagrangian around
the constant vacuum expectation value $\phi_0$ ($\del_{\mu}\phi_0=0$)
in the Riemann normal coordinate. 
Since the metric $g_{ij}(\phi_0)$ is diagonal in this coordinate, we
denote it as ${1\over 2}f_{\pi}^2\delta_{ij}$. We find~\cite{AFM}
\beq
  {\cal L}(\phi_0+\pi)
 &=& \1{2} g_{ij}(\phi_0+\pi)
     \del_{\mu} (\phi_0+\pi)^i \del^{\mu} (\phi_0+\pi)^j \non
\hs{-7} &=& \1{2} f_{\pi}^2 \delta_{ij} \del_{\mu} \xi^i \del^{\mu}\xi^j
  + \1{6} R_{iklj}(\phi_0) \xi^k\xi^l \del_{\mu} \xi^i \del^{\mu}\xi^j
  + O(\xi^5),
\label{ex.of L}
\eeq
where terms up to the fourth order in $\xi^i$ are kept because
they are sufficient in order to compute two-body scattering amplitudes
(figure 2). The fluctuating field $\xi^i$ is interpreted as the particle
propagating on the vacuum $\phi_0$.
To calculate many-body scattering amplitudes,
we must expand the Lagrangian to higher orders in $\xi^i$.
\begin{figure}
 \epsfxsize=5cm
 \centerline{\epsfbox{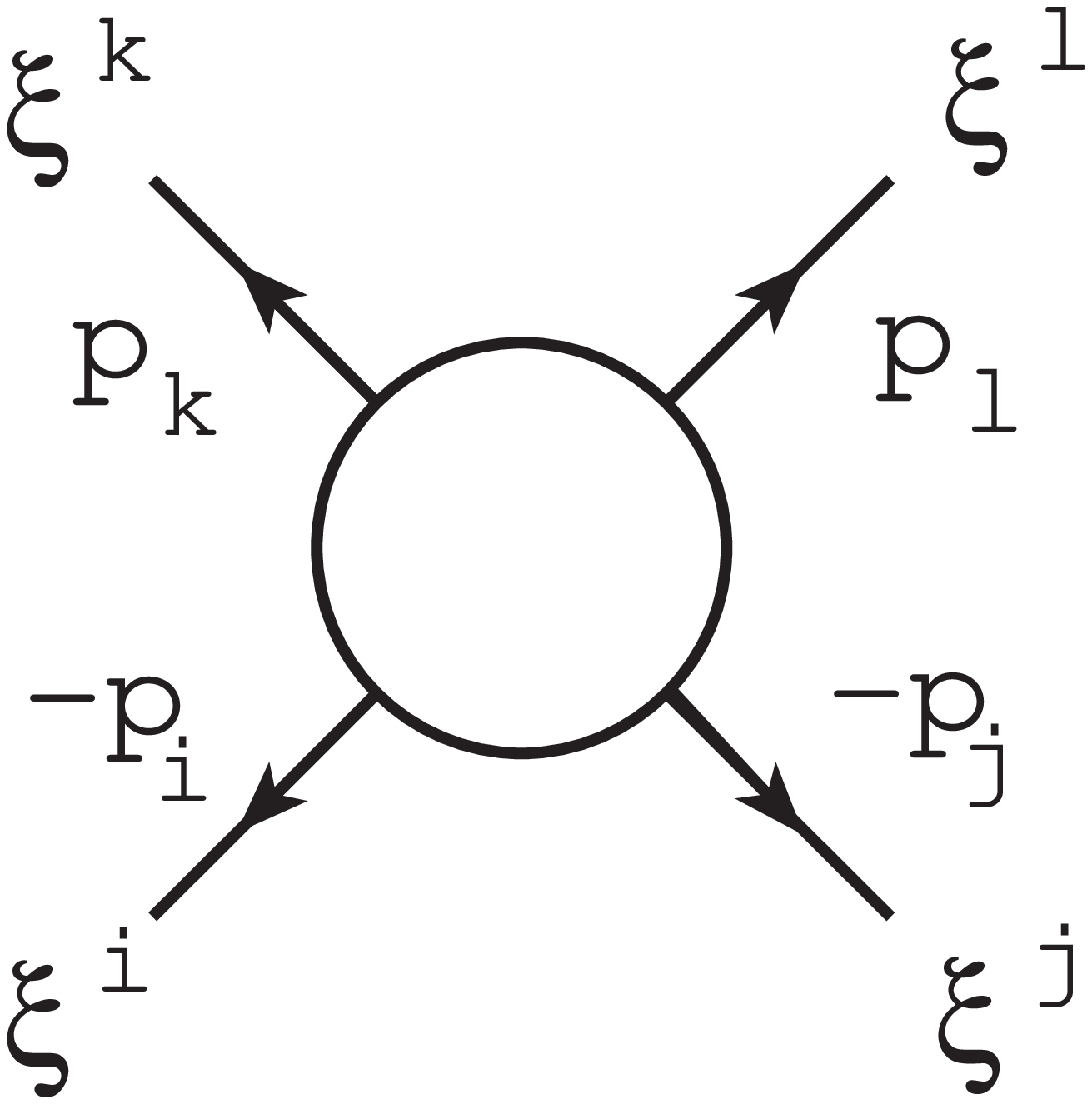}}
 \centerline{\bf Figure\,2.\,Two-body scattering of NG bosons}
\end{figure}
Using the definition
\beq
&& <\xi^k(p_k),\xi^l(p_l)|i {\cal L}_{int} |\xi^i(p_i),\xi^j(p_j)> \non
& &= i(2\pi)^4 \delta^{(4)}(p_k+p_l-p_i-p_j)
   {\cal M}(\xi^i(p_i),\xi^j(p_j) \to \xi^k(p_k),\xi^l(p_l)) ,
\eeq
we find that the two-body scattering amplitudes are in general
expressed in terms of the curvature tensor as
\footnote{Since two point function of $\xi$ is proportional to $f_{\pi}^{-2}$,
we have to supply $f_{\pi}^{-4}$ to obtain the scattering amplitude
properly normalized.}
\beq
&&{\cal M}(\xi^i(p_i),\xi^j(p_j) \to \xi^k(p_k),\xi^l(p_l)) \non
&&= -\1{3f_{\pi}^4}[(s-u)R_{kijl} + (u-t)R_{kjil} + (t-s)R_{kjli}] ,
 \label{scat.amp.}
\eeq
where the Madelstam variables $s,t$ and $u$ are defined by
\beq
 &&s \defeq (p_i + p_j)^2 = +2 p_i \cdot p_j = +2 p_l \cdot p_k, \non
 &&t \defeq (p_i - p_k)^2 = -2 p_i \cdot p_k = -2 p_l \cdot p_j, \non
 &&u \defeq (p_i - p_l)^2 = -2 p_i \cdot p_l = -2 p_j \cdot p_l.
\eeq

Our task is now to determine the metric for the target manifold of
the effective Lagrangian of NG bosons by the symmetries $G$ and $H$.
The coset manifold $G/H$ is parametrized by the NG bosons $\phi^{\alpha}$.
We denote the broken and unbroken Hermitian generators by\footnote{
The Lie algebras of the groups $G$ and $H$ are denoted
by ${\cal G}$ and ${\cal H}$, respectively.
We use the indices $i, j, \cdots$ for the broken generators and
$a, b, \cdots$ for the unbroken generators.}
\beq
 X_i \in {\cal G} - {\cal H} \;,\; H_a \in {\cal H} .
\eeq
The fundamental variable in the construction of the metric is the
representative of the coset $G/H$
\beq
 U(\phi) = e^{i \phi \cdot X} \in G/H
 \;,\; \phi \cdot X =  \phi^{\alpha} X_i \delta^i_{\alpha},
\label{ccwz}
\eeq
which is unitary.

In order to express the metric in terms of the fundamental variable,
it is useful to introduce the Maurer-Cartan 1-form
\beq
 \1{i}U^{-1} dU
 = (e^i_{\alpha}(\phi ) X_i + \omega^a_{\alpha}(\phi ) H_a) d\phi^{\alpha},
 \label{MC1form1}
\eeq
which is a Lie algebra valued 1-form.
The broken generator element of the Maurer-Cartan 1-form,
$e^i_{\alpha}(\phi)$ is called the vielbein,
which connects the metric of target manifold
and the tangent space flat metric.
The unbroken element $\omega^a_{\alpha}(\phi)$
is called the canonical $H$ connection. Using the definition
given in~(\ref{MC1form1}), we can explicitly write the metric of
the target manifold
\beq
 g_{\alpha\beta}(\phi)
 = f_{\pi}^2 e^i_{\alpha}(\phi) e^j_{\beta}(\phi) \delta_{ij} ,
\eeq
by the decay constant of the pion (NG boson) $f_{\pi}$,
which has unit mass dimension and
represents the size of the coset manifold $G/H$.
If $G/H$ is reducible, there are as many decay constants as
the number of the $H$ irreducible factors;
if it is irreducible, there exists only one decay constant.

The Maurer-Cartan's structure equation
\beq
 d(U^{-1} dU) + (U^{-1}dU) \wedge (U^{-1}dU) = 0,
\eeq
enables us to calculate the curvature tensor of the symmetric
irreducible coset manifold $G/H$ as~\cite{KS}
\beq
 R_{ijkl} = f_{\pi}^2 \; {f_{ij}}^a f_{akl}.
 \label{curv.of irre.sym.}
\eeq
The low energy theorems for the two-body scattering amplitudes
of NG bosons then take the form
\beq
 &&{\cal M}(\xi^i(p_i),\xi^j(p_j) \to \xi^k(p_k),\xi^l(p_l)) \non
 &&= -\1{3f_{\pi}^2}[(s-u){f_{ki}}^a f_{ajl} + (u-t){f_{kl}}^a f_{aij}
                   + (t-s){f_{kj}}^a f_{ali} ].
     \label{LET non-SUSY}
\eeq
This is a well-known result in the current algebra~\cite{BKY}.
On the other hand, eq.~(\ref{scat.amp.}) is valid in more general cases.
If the target manifold $G/H$ is non-symmetric or reducible,
the curvature tensor is more complicated~\cite{KS}
and so are the low energy theorems.

In conclusion, the scattering amplitudes are written by
the geometrical quantities (for example, the curvature tensor
in the case of the two-body scattering amplitudes).
The symmetries restrict the form of the metric tensor
of the target manifold of the sigma model and are strong enough
to determine the geometric objects (curvature tensor) in terms of
group structure constants up to an overall factor.

In the next section, we introduce supersymetry in addition to the
global symmetries $G$ and $H$.
At first glance, the restriction seems to be stronger and
the low energy theorems more restrictive. We will see that this is
not true because supersymmetry requires extra massless bosons and
the target manifolds become more complicated.

\section{Scattering amplitudes of NG and QNG bosons in supersymmetric theory}

In $N=1$ supersymmetric theories,
the general Lagrangian for chiral fileds is
\beq
 {\cal L} &=& \int d^2\th d^2\thb K(\Ph,\Phd) \non
 &=& g_{ij^*}(\ph,\ph^*)\del_{\mu}\ph^i \del^{\mu}\ph^{*j}
 + ig_{ij^*}\psb^j \sigb^{\mu}
   (\del_{\mu}\ps^i + {\Gamma^i}_{lk}\del_{\mu}\ph^l \ps^k)\non
 && + \1{4} R_{ij^*kl^*} \ps^i\ps^k \psb^j\psb^l,
\eeq
where the metric tensor of the target K\"{a}hler manifold~\cite{WB} is
given by the K\"{a}hler potential $K$ as\footnote{Here we use the indices
$i,j,\cdots$ for general complex coordinates of the K\"{a}hler target
manifold. They should not be confused with
the indices of broken generators in non-supersymmetric case.}
\beq
 g_{ij^*}(\ph,\ph^*)
 = {\del^2 K(\ph,\ph^*) \over \del \ph^i \del \ph^{*j}}.
\eeq

The chiral superfield has the component fields
\beq
 \Ph &=& \ph + \sqrt{2}\th\ps + \th\th F .
\eeq
The bosonic part is a complex scalar field
\beq
 \ph = A + i B ,
\label{com}
\eeq
which is the coordinate of the target manifold.
Here $A$ bosons always represent the ordinary NG bosons,
but $B$ bosons may be NG bosons or QNG bosons.
In the former case the chiral superfields are called the pure type or
the non-doubled type; in the latter case
they are called the mixed type or the doubled type.
The maximal realization is the nonlinear realization in which
$A$ bosons are NG bosons but all $B$ bosons are QNG bosons.
In this paper we focus on this maximal realization because
it is the most natural one in Lagrangian field theories.
Although we do not discuss the fermions in this paper,
fermions in the chiral superfields
are sometimes called the quasi Nambu-Goldstone fermions.

For the low-energy effective Lagrangian
with the symmetry $G$ broken down to $H$,
the K\"{a}hler potential must have $G$-isometry invariance
(up to the K\"{a}hler transformation).
BKMU constructed $G$-invariant K\"{a}hler potential
by generalizing the CCWZ method.
The broken and unbroken generators of the complexified group are written as
\beq
 Z_R \in {\cal G}^{\bf C} - \hat{\cal H} \;,\; K_M \in \hat{\cal H} .
\eeq
The commutators of these generators are given by
\beq
 \left[ K_M, K_N \right] = i {f_{MN}}^L K_L\;,\;
 \left[ Z_R, K_M \right] = i {f_{RM}}^N K_N\;,\;
 \left[ Z_R, Z_S \right] = i {f_{RS}}^T Z_T ,
\eeq
where in the last equation, we have assumed that the complex coset
$G^{\bf C}/\hat H$ is a symmetric space with an involutive automorphism
\beq
 Z_R \to - Z_R \;,\; K_M \to K_M .  \label{auto mor.}
\eeq
In general, these generators are written by complex linear combinations of
the Hermitian generators of $G$.
The Hermitian generators in ${\cal G}^{\bf C} - \hat{\cal H}$
correspond to the mixed type chiral superfileds,
while the sets of non-Hermitian step generators in
${\cal G}^{\bf C} - \hat{\cal H}$ and $\hat{\cal H}$
correspond to the pure type ones.

A novel feature in the supersymmetric theories is that the
BKMU fundamental variable
\beq
 \xi(\Ph) = {\rm e}^{i \Ph \cdot Z} \in G^{\bf C}/\hat H
 \;,\; \Phi \cdot Z = \sum_{i=1}^{N_{\Phi}} \Phi^i Z_R \delta^R_i,
\label{bkmu}
\eeq
is not unitary. This is due to the complex coordinates (\ref{com}).
This variable is the representative of the complex coset
$G^{\bf C}/\hat H$. The transformation law of $\xi$ is
\beq
 \xi \stackrel{g}{\to} \xi ^{\prime} = g \xi \hat h^{-1}(g,\xi).
\eeq

In order to write down K\"{a}hler potential, let us now
prepare some representation $(\rho,V)$ of $G$.
In the case of low-energy effective Lagrangian of linear model origin,
there are some $\hat{H}$-invariant vectors $v_a,v_b, \cdots \in V$
in the representation space:
\beq
 \rho(\hat H)v_a = v_a.
\eeq
The $G$-invariant K\"{a}hler potential (BKMU A-type Lagrangian) is then
an arbitrary function of $G$-invariant quantities~\cite{Le,BKMU}
\beq
 K(\Ph,\Phd)
= f(v^{\dagger}_a \rho(\xi^{\dagger}(\Ph \dagg)\xi(\Ph))v_b) .
\label{A type}
\eeq
For simplicity, we assume that there is only one $\hat{H}$ invariant vector
$v$ in the representation space.
In this case the coset space $G/H$ is a symmetric one~\cite{Le,KS}.
This assumption is not so restrictive:
For example, this admits the case of the chiral symmetry breaking,
where $G = G_L \times G_R$ breaks down to $H = G_V$ and the full target
manifold is $G^{\bf C}_L \times G^{\bf C}_R /G^{\bf C}_V$~\cite{BKMU}.

In supersymmetric theories, it is useful to use the complex coset
$G^{\bf C}/\hat H$ version of the Maurer-Cartan 1-form (\ref{MC1form1}):
\beq
 \1{i}\xi(\ph)^{-1} d\xi(\ph)
 = ( E^R_i(\ph) Z_R + W^M_i(\ph) K_M ) d\ph^i . \label{MC1form2}
\eeq
The broken generator element of the Maurer-Cartan 1-form $E^R_i(\ph)$ is
called the holomorphic vielbein~\cite{KS}.
This is different from the ordinary non-holomorphic vielbein
connecting the K\"{a}hler and tangent-space flat metrics.
The unbroken generator element $W^M_i(\ph)$ is
called the canonical $\hat H$ connection,
which, together with a K\"{a}hler potential, fixes the spin connection
in the holomorphic formalism~\cite{BL,KS}.
These can be written in the coordinate $\ph$ explicitly as
\beq
 E^R_i(\ph)
 &=& i \delta_i^S
     {\left({1- e^{i\ph \cdot T} \over \ph \cdot T}\right)_S}^R \non
 &=& \delta_i^S
     {\left(1 + \1{2!}i\ph \cdot T + \1{3!}(i\ph \cdot T)^2
                  + \cdots\right)_S}^R ,\\
 W^M_i(\ph)
 &=& i \delta_i^S
     {\left({1- e^{i\ph \cdot T} \over \ph \cdot T}\right)_S}^M \non
 &=& \delta_i^S
  {\left( 1 + \1{2!}i\ph \cdot T + \1{3!}(i\ph \cdot T)^2
      + \cdots\right)_S}^M .
\eeq
Here $T$ is the generator of the adjoint representation of ${\cal G}$;
${(T_C)_A}^B = i {f_{AC}}^B$. We can always choose the point $\ph=0$
to be a symmetric point, where all QNG bosons are set to zero,
since any symmetric points are equivalent under the $G$ isometry.
At this point $\ph=0$, these reduce to
\beq
 E^R_i|_{\ph=0} &=& \delta^R_i ,\label{E on ph=0} \\
 W^M_i|_{\ph=0} &=& 0  .
\label{W on ph=0}
\eeq

For later convenience, let us calculate the derivatives of these at
the point $\ph=0$:
\beq
 \del_j E^R_i|_{\ph=0} &=& -\1{2} \delta_i^S \delta_j^T {f_{ST}}^R
 = 0 \label{d.of E on ph=0}  ,\\
 \del_j W^M_i|_{\ph=0} &=& -\1{2} \delta_i^S \delta_j^T {f_{ST}}^M .
 \label{d.of W on ph=0}
\eeq
In the first equation, we have assumed that $G^{\bf C}/\hat H$ is symmetric.

To calculate geometrical quantities in the holomorphic formalism,
the following formulas resulting from (\ref{MC1form2}) are useful:
\beq
 {\del \xi \over \del \ph^i} v &=& i \xi E^R_i Z_R v,  \non
 {\del \xi \over \del \ph^i} Z_S v
 &=& i \xi (E^R_i Z_R Z_S + W^M_i \left[K_M ,Z_S\right])v \non
 &=& i \xi (E^R_i Z_R Z_S + i W^M_i {f_{MS}}^T Z_T)v  .\label{formula}
\eeq
Using these formulas, we calculate the metric tensor from
(\ref{A type})
\beq
 g_{ij^*}(\ph,\ph^*) = G_{RS^*}(\ph,\ph^*) E^R_i(\ph) (E^S_j(\ph))^*,
  \label{metric}
\eeq
where
\beq
 G_{RS^*}
 = f\pri(v\dagg\xi\dagg\xi v)(v\dagg Z_S\dagg \xi\dagg\xi Z_Rv)
  +f^{\prime\prime}(v\dagg \xi\dagg\xi v) (v\dagg Z_S\dagg \xi\dagg \xi v)
                   (v\dagg \xi\dagg \xi Z_R v),
\label{au}
\eeq
is an auxiliary metric~\cite{KS}. Note that this is not
equal to the Hermitian metric of the (flat) tangent space.

The conventional $G/H$ sigma model is embedded in the full manifold
$G^{\bf C}/\hat H$~\cite{KS}, and the coordinates are spanned by QNG as
well as NG bosons in the form of (\ref{com}). If there are some pure type
multiplets, the corresponding $B$ fields are NG bosons and the BKMU
(\ref{bkmu}) and CCWZ (\ref{ccwz}) variables do not coincide even if we put
all QNG bosons to zero; we need a local $\hat H$ transformation from the
right to compensate the difference~\cite{IKK,BL,KS}. 
In the maximal realization 
we are considering, however, the imaginary directions are all given by QNG
bosons and so the CCWZ (\ref{ccwz}) and BKMU (\ref{bkmu}) variables are
simply connected by
\beq
 \xi(\ph)|_{\rm QNG=0} = U(\phi).
\label{rel.of U-xi}
\eeq
This means that at the symmetric point, $\xi\dagg \xi =1$ and (\ref{au})
reduces to
\beq
 G_{RS^*}^0
 = f\pri(v\dagg v) (v\dagg Z_S\dagg Z_Rv)
 + f^{\prime\prime}(v\dagg v)(v\dagg Z_S\dagg v)(v\dagg Z_R v),
\eeq
where the superscript 0 stands for quantities at the symmetric point (QNG=0).

The symmetric space $G^{\bf C}/\hat H$ is characterized by the involutive
automorphism (\ref{auto mor.}). Since $\hat H$ invariant vector $v$
can be chosen as an eigenvector of the involution with eigenvalues $\pm 1$,
$v\dagg Z_R v$ changes sign under the involution and hence vanishes.
By a suitable choice of broken generators, $Z_S v$ becomes orthonormal
basis of the representation space and we have
$v\dagg Z_R\dagg Z_S v = \delta_{R^*S}$.
The auxiliary metric at the symmetric point then becomes
\beq
G_{RS^*}^0 &=& f\pri(v\dagg v) v\dagg v \; \delta_{RS^*} \non
&=& \1{2} f_{\pi}^2 \delta_{RS^*},
\label{auxi.metric on sym.point}
\eeq
where we have defined the decay constant of the pion (NG boson) by
\beq
 f_{\pi}^2 \defeq 2 f\pri(v\dagg v) v\dagg v  \label{rel of fpi and f},
\label{fpi}
\eeq
which characterizes the size of the compact submanifold $G/H$.
The normalization in (\ref{fpi}) is fixed so as to
give canonically normalized kinetic terms for NG bosons.

{}For the maximal realization of our interest,
the complex version of the Maurer-Cartan form (\ref{MC1form2})
becomes the conventional one (\ref{MC1form1})
without any local $\hat H$ transformation
because of the relation (\ref{rel.of U-xi}).
So the holomorphic vielbein becomes the ordinary vielbein on $G/H$,
and the effective Lagrangian for $A$ bosons takes
the same form as the bosonic case:
\beq
 {\cal L}_{\rm boson}|_{\rm QNG=0}
= \1{2}f_{\pi}^2 e^i_{\alpha} e^j_{\beta} \delta_{ij}
  \del_{\mu}A^{\alpha} \del^{\mu}A^{\beta}
= \1{2}g_{\alpha\beta} \del_{\mu}A^{\alpha} \del^{\mu}A^{\beta} .
\eeq
The fact that we recover the conventional $G/H$ model
by putting QNG$=0$ in the effective Lagrangian can be
understood naturally as follows:
If supersymmetry is broken,
the QNG bosons get the mass of the supersymmetry breaking scale.
We can then integrate out the massive QNG bosons,
and obtain the conventional $G/H$ model, making the curvature tensor of
the obtained manifold identical to eq. (\ref{curv.of irre.sym.}).
The low energy theorems for the scattering amplitudes
of NG bosons are the same as eq. (\ref{LET non-SUSY}).

~From now on we concentrate on the geometric quantities at the symmetric
point. We apply these to the low-energy scattering amplitudes of NG and
QNG bosons. Let us first show that the coset coordinate $\ph$ is
a special one called the adapted coordinate.
According to eqs.~(\ref{E on ph=0}) and (\ref{auxi.metric on sym.point}),
the ordinary metric at the point $\ph=0$ becomes
\beq
 g_{ij^*}|_{\ph=0} =  G_{RS^*}^0 E^R_i|_{\ph=0} (E^S_j)^*|_{\ph^*=0}
 =  \1{2} f_{\pi}^2 \delta_{ij^*}.  \label{metric on ph=0}
\eeq
The derivative of the auxiliary metric is
\beq
\del_j G_{RS^*}
   &=& i \left\{ f\pri(z)(v\dagg Z_S\dagg \xi\dagg\xi Z_T Z_R v)
            +f\prip(z)(v\dagg \xi\dagg\xi Z_T v)
             (v\dagg Z_S\dagg \xi\dagg\xi Z_R v) \right. \non
   &&\; + f\prip(z)(v\dagg Z_S\dagg \xi\dagg\xi Z_T v)
                (v\dagg \xi\dagg\xi Z_R v)
        + f\prip(z)(v\dagg Z_S\dagg \xi\dagg\xi v)
                (v\dagg \xi\dagg\xi Z_T Z_R v)            \non
   &&\; \left.+ f\pripp(z) (v\dagg \xi\dagg\xi Z_T v)
                 (v\dagg Z_S\dagg\xi\dagg\xi v)
                 (v\dagg \xi\dagg\xi Z_R v) \right\} E^T_j \non
&& \hs{-7} + i\left\{ f\pri(z)(v\dagg Z_S\dagg \xi\dagg\xi Z_T v)
     + f\prip(z)(v\dagg Z_S\dagg \xi\dagg\xi v)
                (v\dagg \xi\dagg\xi Z_T v) \right\} i{f_{MR}}^T W^M_j,
\label{d.of G}
\eeq
where $z=v\dagg \xi\dagg \xi v$.
The derivative of the metric (\ref{metric}) is
\beq
 \del_j g_{kl^*} = (\del_j G_{RS^*})E^R_k (E^S_l)^*
  + G_{RS^*}(\del_j E^R_k) (E^S_l)^* .
\eeq
At the symmetric point, this reduces to
\beq
 {\del_j g_{kl^*}} ^0
=  \1{2} f_{\pi}^2 (- f_{MRS^*} W^M_j E^R_k (E^S_l)^*
  + \delta_{RS^*} (\del_j E^R_k) (E^S_l)^*)^0,
\eeq
and especially at the point $\ph=0$ to
\beq
 \del_j g_{kl^*}|_{\ph=0} = 0   \label{d.of g on ph=0},
\eeq
as a result of eqs. (\ref{W on ph=0}) and
(\ref{d.of E on ph=0}) when $G^{\bf C}/\hat H$ is symmetric.
Since the inverse metric at the symmetric point is
\beq
 {g^{ij^*}}^0 = 2 f_{\pi}^{-2} \delta^{RS^*} (E^i_R (E^j_S)^*)^0 ,
\eeq
the (coordinate) connection at the symmetric point is given by
\beq
 {{\Gamma^i}_{jk}}^0
 &=& (g^{il^*}\del_j g_{kl^*})^0 \non
 &=& (- {f_{MR}}^S W^M_j E^R_k E_S^i + (\del_j E^S_k ) E^i_S)^0 ,
\eeq
and especially on the point $\ph=0$ by
\beq
 {\Gamma^i}_{jk}|_{\ph=0} = 0  \label{connection on ph=0}
\eeq
due to eqs. (\ref{W on ph=0}) and
(\ref{d.of E on ph=0}) when $G^{\bf C}/\hat H$ is symmetric.
The coordinate with the properties (\ref{metric on ph=0}) and
(\ref{connection on ph=0}) is called the adapted coordinate on the point
$\ph=0$.\footnote{
That we can choose an adapted coordinate on any point
is a necessary and sufficient condition
that a complex manifold should be a K\"{a}hler manifold.}

Because of eqs. (\ref{d.of g on ph=0}), (\ref{d.of E on ph=0})
and (\ref{E on ph=0}),
the curvature tensor in the adapted coordinate becomes
\beq
 R_{ij^*kl^*}|_{\ph=0} &=& g_{ij^*,kl^*} |_{\ph=0}
 = {\del^4 K \over \del \ph^i \del \ph^{*j} \del \ph^k \del \ph^{*l}}
   |_{\ph=0} \non
 &=& (\del_k \del_{l^*} G_{RS^*})|_{\ph=0} \delta^R_i (\delta^S_j)^*  .
\eeq
According to eq.~(\ref{d.of G}),
the second derivative of the auxiliary metric is
\beq
(\del_k \del_{l^*} G_{RS^*})|_{\ph=0}
 &=& [f\pri(v\dagg v) (v\dagg Z_S\dagg Z_V\dagg Z_U Z_R v)\non
 && + g^2 (\delta_{RU} \delta_{S^*V^*} + \delta_{RV^*} \delta_{US^*}
            +\delta_{RS^*} \delta_{UV^*})]\delta^U_k (\delta^V_l)^*,
\eeq
and the curvature tensor at the point $\ph=0$ in the adapted coordinate is
\beq
R_{ij^*kl^*}|_{\ph=0}
 &&= [f\pri(v\dagg v) (v\dagg Z_S\dagg Z_V\dagg Z_U Z_R v)
   + g^2 (\delta_{RU} \delta_{S^*V^*} + \delta_{RV^*} \delta_{US^*}
            +\delta_{RS^*} \delta_{UV^*})] \non
  &&\;\;\;\;\;
   \times \delta^R_i (\delta^S_j)^* \delta^U_k (\delta^V_l)^* ,
\label{curv.Kahler}
\eeq
where we have defined the constant $g$ by
\beq
 g^2 \defeq f^{\prime\prime}(v\dagg v) (v\dagg v)^2  .
\eeq
The constant $f_{\pi}$ determines the total size of
the compact submanifold $G/H$. On the other hand, the constant $g$
and higher derivatives of the arbitrary function $f$
control the shape of the non-compact directions of the target manifold.
Note that the curvature (\ref{curv.Kahler}) has
the symmetry of K\"{a}hler manifold $R_{ij^*kl^*} = R_{kj^*il^*}$,
since $R_{ij^*kl^*} - R_{kj^*il^*}$ contains the commutators of broken
generators, which produce unbroken generators when $G^{\bf C}/\hat H$ is
symmetric, and vanishes upon multiplying the $\hat H$ invariant vector $v$.

In sect.~2, we have found that the scattering amplitudes
are written in terms of the curvature tensors of the target manifold.
Let us repeat the same analysis here with the above formalism.
Supersymmetry requires that the target manifold should be K\"{a}hler.
This requirement restricts the form of the curvature tensor
of the target manifold and also the low energy theorems.
The low-energy effective Lagrangian is expanded by the
fields $\ph$ and $\psi$ around the point $\ph=0$ up to fourth terms as
\beq
 {\cal L}
 &=& g_{ij^*}|_{\ph=0}\; \del_{\mu}\ph^i \del^{\mu}\ph^{*j}
 + i g_{ij^*}|_{\ph=0}\; \psb^j \sigb^{\mu} \del_{\mu}\ps^i  \non
 &+& R_{ij^*kl^*}|_{\ph=0}\; \ph^k \ph^{*l} \del_{\mu}\ph^i \del^{\mu}\ph^{*j}
   + \1{4} R_{ij^*kl^*}|_{\ph=0}\; \ps^i\ps^k \psb^j\psb^l \non
 &+& i R_{ij^*kl^*}|_{\ph=0}\;
     \ph^{*j}\del_{\mu}\ph^i (\psb^l \sigb^{\mu} \ps^k ) .
\eeq
In deriving this result, we have made a holomorphic coordinate
transformation to eliminate non-covariant terms.
The last three terms are the interaction Lagrangians,
which are described by the curvature tensor of the K\"{a}hler manifold.
They give the two-body scattering amplitudes by the curvature.

The low-energy scattering amplitudes are directly connected with
the curvature tensor in the real basis. In such basis,
the components of the curvature tensor
in the K\"{a}hler manifold are divided into four types
\beq
 \cases{
     R_{A^iA^jA^kA^l} = R_{B^iB^jB^kB^l}
  =  R_{A^iA^jB^kB^l} = R_{B^iB^jA^kA^l} \cr
     R_{B^iA^jB^kA^l} = R_{A^iB^jA^kB^l}
  =- R_{B^iA^jA^kB^l} = -R_{A^iB^jB^kA^l} \cr
     R_{B^iA^jA^kA^l} = - R_{A^iB^jA^kA^l}
 = - R_{A^iB^jB^kB^l} = R_{B^iA^jB^kB^l} \cr
     R_{A^iA^jB^kA^l} = - R_{A^iA^jA^kB^l}
 = - R_{B^iB^jA^kB^l} = R_{B^iB^jB^kA^l} \cr},
\label{sym.of R}
\eeq
as a result of the K\"{a}hler conditions.\footnote{
These are equivalent to the  vanishing curvature conditions
$R_{ijkl}=R_{i^*jkl}=R_{ij^*kl}=R_{ijk^*l}=R_{ijkl^*}=R_{i^*j^*kl}=0$
and their Hermitian conjugates.}
These curvatures have the symmetry under
\beq
 \cases { A^i \to + B^i \cr B^i \to - A^i} ,
\eeq
and so do the scattering amplitudes.

The first equation in (\ref{sym.of R}) implies
that the two-body amplitudes of the QNG bosons
coincide with those of the corresponding NG bosons
\beq
 {\cal M}({\rm NG,NG \to NG,NG}) = {\cal M}({\rm QNG,QNG \to QNG,QNG}),
\eeq
owing to eq.~(\ref{scat.amp.}) (see figure 3).
Note that this is a result following solely from the K\"{a}hler conditions
and is valid for general K\"{a}hler sigma models.

\begin{figure}
 \epsfxsize=8cm
 \centerline{\epsfbox{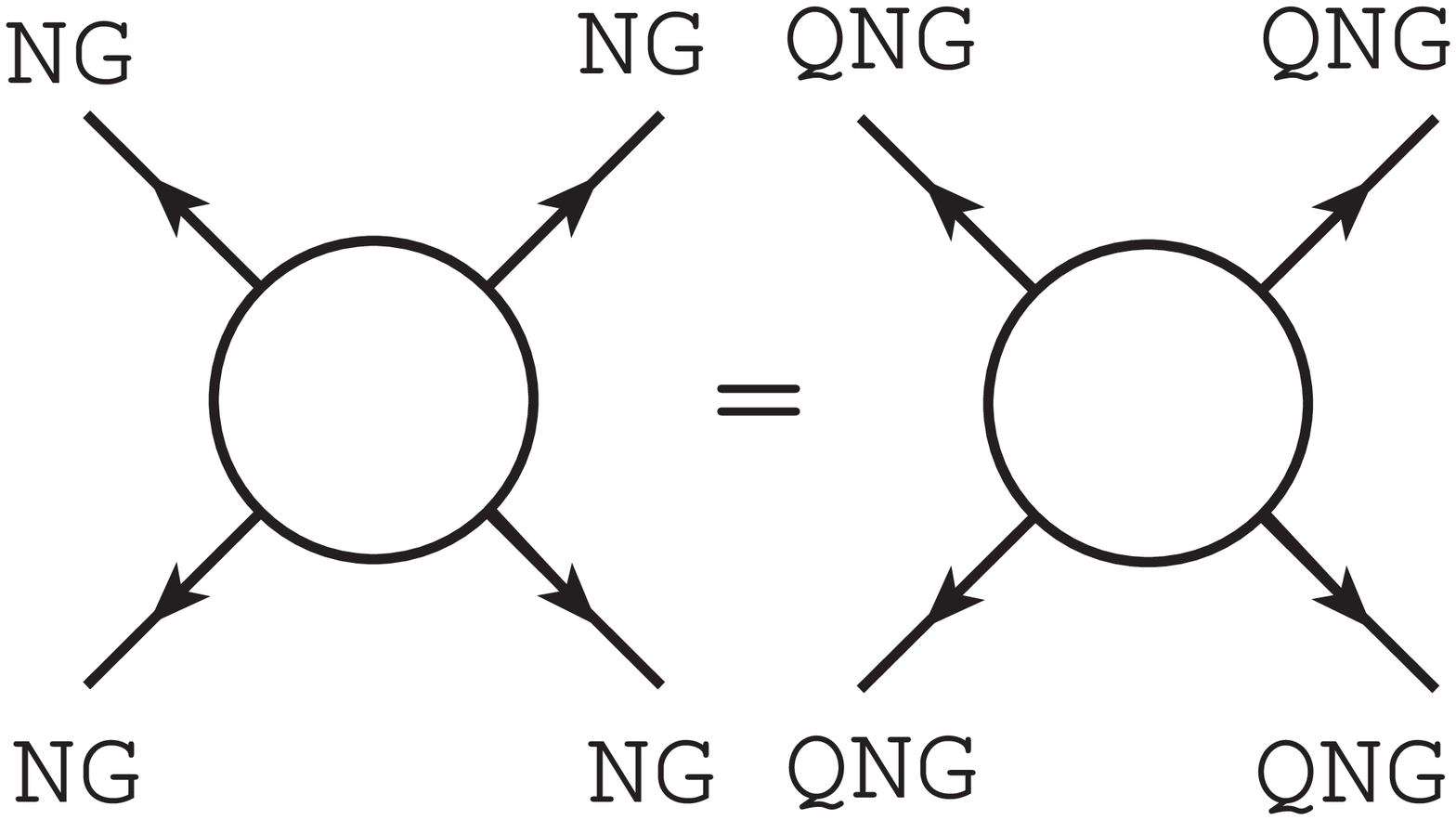}}
 \centerline{\bf Figure\,3.\,Scattering amplitudes of QNG
                             coincide with those of NG}
\end{figure}

Let us now evaluate the actual values of the curvature tensors in
the real basis. By using (\ref{curv.Kahler}) and
\beq
 \del_{A^i} = \del_i + \del_i^* \;,\;
 \del_{B^i} = i(\del_i - \del_i^*) ,
\eeq
we derive\footnote{For notational simplicity, we do not write $\ph=0$
on the right hand side.}
\beq
 R_{A^i A^j A^k A^l}|_{\ph=0}
 &=& R_{ij^*kl^*} + R_{i^*jkl^*} + R_{ij^*k^*l} + R_{i^*jk^*l} \non
 &=& R_{ij^*kl^*} - R_{ji^*kl^*} - R_{ij^*lk^*} + R_{ji^*lk^*} \non
 &=& f\pri (v\dagg v)v\dagg (Z_S [Z_V,Z_U] Z_R - Z_R [Z_V,Z_U] Z_S)v
     \delta^R_i \delta^S_j \delta^U_k \delta^V_l\non
 &=& f_{\pi}^2 {f_{RS}}^M f_{MUV}
  \delta^R_i \delta^S_j \delta^U_k \delta^V_l ,
  \label{R1}
\eeq
where we have also used eq.~(\ref{rel of fpi and f}).
This depends only on the first derivative of the arbitrary function
and the higher derivatives all cancel out.
It is thus completely independent of the shape of the non-compact directions.
This result demonstrates that the curvature tensors and hence the low energy
theorems for the NG bosons coincide with those in non-supersymmetric theories.

Similarly, we can obtain the components of the curvature tensor involving
both the NG and QNG bosons.
One finds
\beq
 R_{B^i A^j B^k A^l}|_{\ph=0}
 &=& i^2(R_{ij^*kl^*} - R_{i^*jkl^*} - R_{ij^*k^*l} + R_{i^*jk^*l}) \non
 &=& - f\pri (v\dagg v) v\dagg
   (Z_S \{Z_V,Z_U\} Z_R + Z_R \{Z_V,Z_U\}Z_S)v
   \delta^R_i \delta^S_j \delta^U_k \delta^V_l \non
 &&  - 4 g^2 (\delta_{ik} \delta_{jl} + \delta_{il} \delta_{kj}
            + \delta_{ij} \delta_{kl} ) .     \label{R2}
\eeq
These components include higher derivatives of the arbitrary function,
which controls the non-compact QNG directions, showing their dependence
on the shape of non-compact directions and the details of the underlying
fundamental theory. The low-energy scattering amplitudes for two NG bosons
and two QNG bosons are calculated by use of eqs.~(\ref{R2}),
(\ref{sym.of R}) and (\ref{scat.amp.}). They also depend on the higher
derivatives of the arbitrary function.

We also find from (\ref{sym.of R})\footnote{
Actually these two quantities in (\ref{cu1}) and (\ref{cu2}) are not
independent since $R_{B^i A^j A^k A^l} = R_{A^k A^l B^i A^j}$.}
\beq
 R_{B^i A^j A^k A^l}|_{\ph=0}
 &=& i(R_{ij^*kl^*} - R_{i^*jkl^*} + R_{ij^*k^*l} - R_{i^*jk^*l}) \non
 &=& if\pri (v\dagg v)v\dagg (Z_S [Z_V,Z_U] Z_R + Z_R [Z_V,Z_U] Z_S)v
     \delta^R_i \delta^S_j \delta^U_k \delta^V_l \non
 &=& 0,
\label{cu1}
\\
 R_{A^i A^j B^k A^l}|_{\ph=0}
 &=& i(R_{ij^*kl^*} + R_{i^*jkl^*} - R_{ij^*k^*l} - R_{i^*jk^*l}) \non
 &=& i f\pri (v\dagg v)
     v\dagg (Z_V [Z_S,Z_R] Z_U + Z_U [Z_S,Z_R] Z_V)
     \delta^R_i \delta^S_j \delta^U_k \delta^V_l \non
 &=& 0.
\label{cu2}
\eeq
{}From these two equations and eq. (\ref{sym.of R}), we find that
all components of the curvature tensor which include odd numbers of
NG bosons and QNG bosons vanish. Eq.~(\ref{scat.amp.}) then tells us that
the low-energy scattering amplitudes with odd numbers of NG bosons
and QNG bosons all vanish.

\section{Conclusions and Discussions}

In supersymmetric theories whose global symmetry $G$ is broken down to
its subgroup $H$, there appear the QNG bosons besides the ordinary NG bosons.
The low-energy effective Lagrangian can be described by a nonlinear sigma
model whose target manifold is a K\"{a}hler coset $G^{\bf C}/\hat H$
with only the compact symmetry $G$. The manifold is non-compact and
its non-compact directions correspond to the QNG bosons.
Since the global symmetry does not restrict the shape of the non-compact
directions, the K\"{a}hler potential of the effective Lagrangian includes
an arbitrary function.

Nevertheless we have shown that the low-energy scattering amplitudes which
has only QNG bosons are equal to those of the corresponding NG bosons.
The arbitrariness comes into the low-energy scattering amplitudes
which has two NG bosons and two QNG bosons.
The low-energy scattering amplitudes which have the odd number of NG bosons
and QNG bosons all vanish for the case of symmetric point.

In this article, we have investigated only the two-body scattering amplitudes.
However, generalization to many-body scattering amplitudes is straightforward.
If we expand the Lagrangian in the Riemann normal coordinate to higher
orders in $\xi$, the coefficients are covariant geometrical objects
such as covariant derivatives of curvature tensor.
We can obtain (exact) low-energy scattering amplitudes
by using the effective Lagrangian.
They will have the symmetry NG $\to$ QNG, QNG $\to-$NG
similar to the two-body amplitudes,
and we can obtain the low energy theorems of NG and QNG bosons.

Another interesting extension of our work is to examine how our results are
generalized at non-symmetric point. We expect that such a generalization
brings new features into the low energy theorems.

\section*{Acknowledgement}

One of the authors (M. Nitta) is very grateful to
S. Iso and N. Ishibashi for useful discussions.

\newpage

\end{document}